\documentclass[reprint,superscriptaddress,aps,prl]{revtex4-1}
\usepackage{amsmath,epsfig,amssymb,subfigure,bm,dsfont}
\usepackage{lipsum} 
\usepackage{color}
\usepackage{graphicx}
\usepackage{dcolumn}
\usepackage{bm}
\usepackage{bbold}
\usepackage{amsfonts,color}
\usepackage{amsmath}
\usepackage{soul}
\newcounter{lastnote}

\begin{document}

\title{Observation of spin-wave moir\'e edge and cavity modes in twisted magnetic lattices}

\author{Hanchen Wang}
\thanks{These authors contributed equally to this work.}
\affiliation{%
Fert Beijing Institute, MIIT Key Laboratory of Spintronics, School of Integrated Circuit Science and Engineering, Beihang University, Beijing 100191, China
}
\affiliation{%
International Quantum Academy, Shenzhen, China
}%
\affiliation{%
Department of Materials, ETH Zurich, Zurich 8093, Switzerland
}%

\author{Marco Madami}
\thanks{These authors contributed equally to this work.}
\email{marco.madami@unipg.it}
\affiliation{%
Dipartimento di Fisica e Geologia, Universit\`a di Perugia, Perugia I-06123, Italy
}

\author{Jilei Chen}
\thanks{These authors contributed equally to this work.}
\affiliation{%
Shenzhen Institute for Quantum Science and Engineering (SIQSE), and Department of Physics,
Southern University of Science and Technology (SUSTech), Shenzhen 518055, China
}
\affiliation{%
International Quantum Academy, Shenzhen, China
}%

\author{Hao Jia}
\thanks{These authors contributed equally to this work.}
\affiliation{%
Shenzhen Institute for Quantum Science and Engineering (SIQSE), and Department of Physics,
Southern University of Science and Technology (SUSTech), Shenzhen 518055, China
}
\affiliation{%
International Quantum Academy, Shenzhen, China
}%

\author{Yu Zhang}
\thanks{These authors contributed equally to this work.}
\affiliation{%
Beijing National Laboratory for Condensed Matter Physics, Institute of Physics, University of Chinese Academy of Sciences, Chinese Academy of Sciences, Beijing, China
}%

\author{Rundong Yuan}
\thanks{These authors contributed equally to this work.}
\affiliation{%
Fert Beijing Institute, MIIT Key Laboratory of Spintronics, School of Integrated Circuit Science and Engineering, Beihang University, Beijing 100191, China
}%

\author{Yizhan~Wang}
\affiliation{%
Beijing National Laboratory for Condensed Matter Physics, Institute of Physics, University of Chinese Academy of Sciences, Chinese Academy of Sciences, Beijing, China
}%

\author{Wenqing He}
\affiliation{%
Beijing National Laboratory for Condensed Matter Physics, Institute of Physics, University of Chinese Academy of Sciences, Chinese Academy of Sciences, Beijing, China
}

\author{Lutong Sheng}
\affiliation{%
Fert Beijing Institute, MIIT Key Laboratory of Spintronics, School of Integrated Circuit Science and Engineering, Beihang University, Beijing 100191, China
}%

\author{Yuelin Zhang}
\affiliation{%
Fert Beijing Institute, MIIT Key Laboratory of Spintronics, School of Integrated Circuit Science and Engineering, Beihang University, Beijing 100191, China
}%

\author{Jinlong Wang}
\affiliation{%
Fert Beijing Institute, MIIT Key Laboratory of Spintronics, School of Integrated Circuit Science and Engineering, Beihang University, Beijing 100191, China
}%

\author{Song Liu}
\affiliation{%
	International Quantum Academy, Shenzhen, China
}%
\affiliation{%
Shenzhen Institute for Quantum Science and Engineering (SIQSE), and Department of Physics,
Southern University of Science and Technology (SUSTech), Shenzhen 518055, China
}

\author{Ka Shen}
\affiliation{%
Department of Physics, Beijing Normal University, Beijing, China
}

\author{Guoqiang~Yu}
\affiliation{%
	Beijing National Laboratory for Condensed Matter Physics, Institute of Physics, University of Chinese Academy of Sciences, Chinese Academy of Sciences, Beijing, China
}%
\author{Xiufeng Han}
\affiliation{%
Beijing National Laboratory for Condensed Matter Physics, Institute of Physics, University of Chinese Academy of Sciences, Chinese Academy of Sciences, Beijing, China
}%

\author{Dapeng Yu}
\affiliation{%
International Quantum Academy, Shenzhen, China
}%
\affiliation{%
Shenzhen Institute for Quantum Science and Engineering (SIQSE), and Department of Physics,
Southern University of Science and Technology (SUSTech), Shenzhen 518055, China
}

\author{Jean-Philippe Ansermet}
\email{jean-philippe.ansermet@epfl.ch}
\affiliation{%
Institute of Physics, Ecole Polytechnique F\'ed\'erale de Lausanne (EPFL), 1015, Lausanne, Switzerland
}
\affiliation{%
	Shenzhen Institute for Quantum Science and Engineering (SIQSE), and Department of Physics,
	Southern University of Science and Technology (SUSTech), Shenzhen 518055, China
}
\author{Gianluca Gubbiotti}
\email{gubbiotti@iom.cnr.it}
\affiliation{%
Istituto Officina dei Materiali del Consiglio Nazionale delle Ricerche (IOM-CNR), c/o Dipartimento di Fisica e Geologia, Perugia, Italy.
}%

\author{Haiming Yu}
\email{haiming.yu@buaa.edu.cn}
\affiliation{%
Fert Beijing Institute, MIIT Key Laboratory of Spintronics, School of Integrated Circuit Science and Engineering, Beihang University, Beijing 100191, China
}
\affiliation{%
International Quantum Academy, Shenzhen, China
}%
\date{\today}

\begin{abstract}
We report the experimental observation of the spin-wave moir\'e edge and cavity modes using Brillouin light scattering spectro-microscopy in a nanostructured magnetic moir\'e lattice consisting of two twisted triangle antidot lattices based on an yttrium iron garnet thin film. Spin-wave moir\'e edge modes are detected at an optimal twist angle and with a selective excitation frequency. At a given twist angle, the magnetic field acts as an additional degree of freedom for tuning the chiral behavior of the magnon edge modes. Micromagnetic simulations indicate that the edge modes emerge within the original magnonic band gap and at the intersection between a mini-flatband and a propagation magnon branch. Our theoretical estimate for the Berry curvature of the magnon-magnon coupling suggests a non-trivial topology for the chiral edge modes and confirms the key role played by the dipolar interaction. Our findings shed light on the topological nature of the magnon edge mode for emergent moir\'e magnonics.
\end{abstract}

\maketitle

\section{I. Introduction}
When two stacking lattices are slightly twisted with one another~\cite{Bistritzer2011} or have a small lattice mismatch~\cite{Tong2017}, a new periodical pattern arises, known as a moir\'e superlattice with a new periodicity significantly larger than the original lattice constant. Moir\'e superlattices comprising twisted layers of van der Waals materials exhibit extraordinary electronic behaviours such as superconductivity and correlated topological states~\cite{Bistritzer2011,Tong2017,Cao2018,Pierce2021,Mak2021,WYao2022}. Magic-angle twisted bilayer graphene~\cite{Bistritzer2011} as a moir\'e superlattice has been intensively investigated owing to its novel electronic properties such as unconventional superconductivity~\cite{Cao2018} and correlated insulating and ferromagnetic phases~\cite{Pierce2021}. The concept of moir\'e physics has recently been extended and applied in photonics to engineer photonic band structures providing novel functionalities such as magic-angle lasers~\cite{Dong2021,Mao2021}, forming novel band structures such as moir\'e flatband or mini-flatband with narrow bandwidths. Magnons or spin waves~\cite{Chumak2015,Han2019,Barman2021,Pirro2021,Zhang2020}, are collective spin excitations that can propagate in magnetic metals~\cite{Han2019} and insulators~\cite{Cornelissen2015} free of charge transport and therefore are applicable for low-power computing devices, such as magnonic logic gates~\cite{Khitun2010,QWang2020}. To date, moir\'e physics in magnonic systems has only been studied theoretically~\cite{Li2020,WangC2020,Chen2022}. Topological spin-wave edge modes have been theoretically predicted in several systems~\cite{Zhang2013,Elyasi2019,Mook2014,Owerre2016,Kim2016,Hiros2020,Kondo2021,WangXG2020,Bos2022}, such as bicomponent magnonic lattices~\cite{Shindou2013} and stitched magnonic lattices~\cite{Li2018}, but up to now have not been experimentally demonstrated in any material systems.
\begin{figure*}
\centering
\includegraphics[width=178mm]{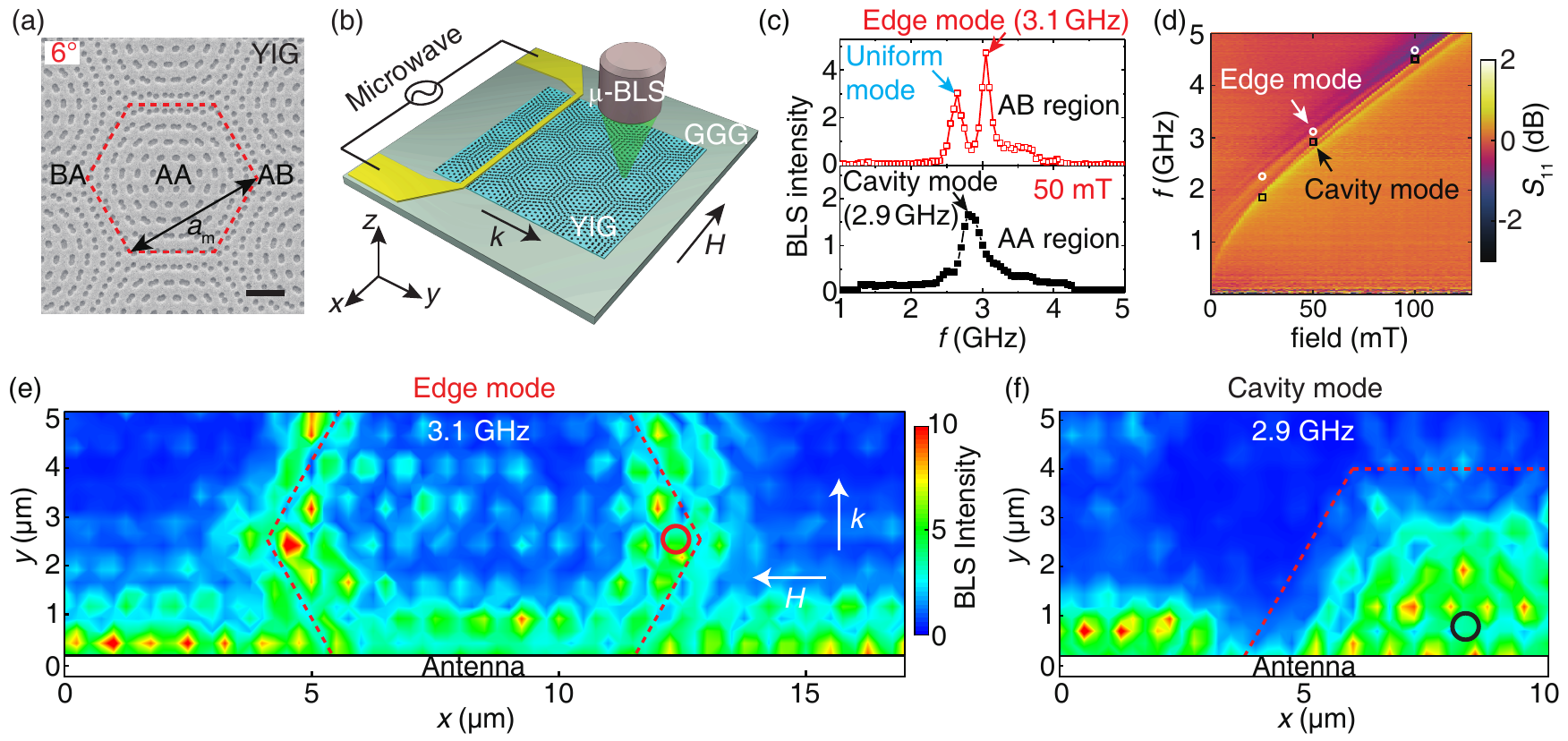}
\caption{(a) Scanning electron microscope (SEM) image of a moir\'e magnonic lattice based on YIG grown on a GGG substrate with a twist angle of 6$^\circ$. The red dashed line indicates a moir\'e unit cell with commensurate AA region at its center and incommensurate AB (BA) region at its edge. Moir\'e lattice constant $a_{\rm m}$ is marked by the black arrow. Scale bar, 2~$\mu$m. (b) Schematics of spatially-resolved spin-wave measurement on moir\'e magnonic lattices based on $\mu$-BLS. Microwaves injected by a nanostripline antenna (gold line) excite spin waves with a wavevector $k$ perpendicular to the antenna. Magnetic field $H$ is applied along the antenna. (c) $\mu$-BLS signals detected as a function of frequency at the AB region [red circle in (e)] and AA region [black circle in (f)] with an applied field of 50~mT. (d) Spin-wave reflection spectra $S_{\rm 11}$ measured by the all-electrical spin-wave spectroscopy. White circles and black open squares: Field dependent edge and cavity mode frequencies detected by $\mu$-BLS, respectively. (e) Two-dimensional spin-wave intensity maps measured by $\mu$-BLS at 3.1~GHz. The center of the excitation antenna is defined as $y$\,=\,0. (f) Two-dimensional spin-wave intensity maps measured by $\mu$-BLS at 2.9~GHz. The red dashed lines are guide to the eyes for a moir\'e unit cell. The applied field is set at 50~mT.}
\label{fig1}
\end{figure*}

In this Letter, we experimentally investigate spin-wave propagation in a moir\'e magnonic lattice and observe topological spin-wave edge states at the boundary of a moir\'e unit cell with an optimal combination of the twist angle and applied magnetic field. Two antidot sublattices with a relative twist angle are merged in a single yttrium iron garnet (YIG) thin film, thus forming a moir\'e magnonic lattice [Fig.~\ref{fig1}(a)]. Here we choose a triangle lattice to resemble the hexagonal lattice symmetry of graphene~\cite{Cao2021,TYu2021} while the results may apply to other types of lattice~\cite{Chen2022}. Antidot lattices~\cite{Neusser2010,Duerr2011,Gross2021} are used to form the moiré magnonic crystals instead of dot arrays~\cite{Tacchi2011} to preserve large continuous film areas for efficient spin-wave guiding. We employ micro-focused Brillouin light scattering ($\mu$-BLS) [Fig.~\ref{fig1}(b)] to directly visualize two types of spin-wave modes in a moir\'e magnonic lattice, namely, i) Spin waves propagating along the edges of a moir\'e unit cell [Fig.~\ref{fig1}(e)], which we refer to henceforth as moir\'e edge modes or simply edge modes. ii) Spin waves strongly confined at the center of a moir\'e unit cell [Fig.~\ref{fig1}(f)], which is referred to as moir\'e cavity modes or simply cavity modes in analogy to its photonic counterpart~\cite{Dong2021,Mao2021}. The most intense edge mode arises in the moir\'e magnonic lattice at an optimal twist angle of 6$^\circ$ with an applied field of 50~mT. Micromagnetic simulations indicate that the edge mode emerges within the original band gap~\cite{XWang2017,Bos2020,Dai2021} and at the intersection between two magnon branches. By extending the theoretical tools in magnon-photon~\cite{Okamoto2020} and magnon-phonon~\cite{Okamoto20202} systems, we derive the Berry curvature induced by the magnon-magnon coupling with dipolar interactions~\cite{Shindou2013} that reveals a nontrival topology of the chiral spin-wave edge modes. The moir\'e edge modes reported here are related but still distinctive from high-energy magnons ($\sim$meV) in van der Waals materials usually studied with large facilities such as neutron scattering~\cite{Yuan2020}, at ultra-low temperatures below 1 K~\cite{Gan_arXiv} and with high magnetic fields typically of a few Tesla~\cite{Dai2021}. In this study, spin-wave edge modes are excited at GHz frequencies, detected at room temperature and tunable with a moderate magnetic field, and are thus readily compatible with existing on-chip electronics and microwave technology providing new opportunities towards applications in spin-wave-based computing~\cite{Chumak2022} and wireless communications~\cite{Harris2022}.
\begin{figure*}
\centering
\includegraphics[width=178mm]{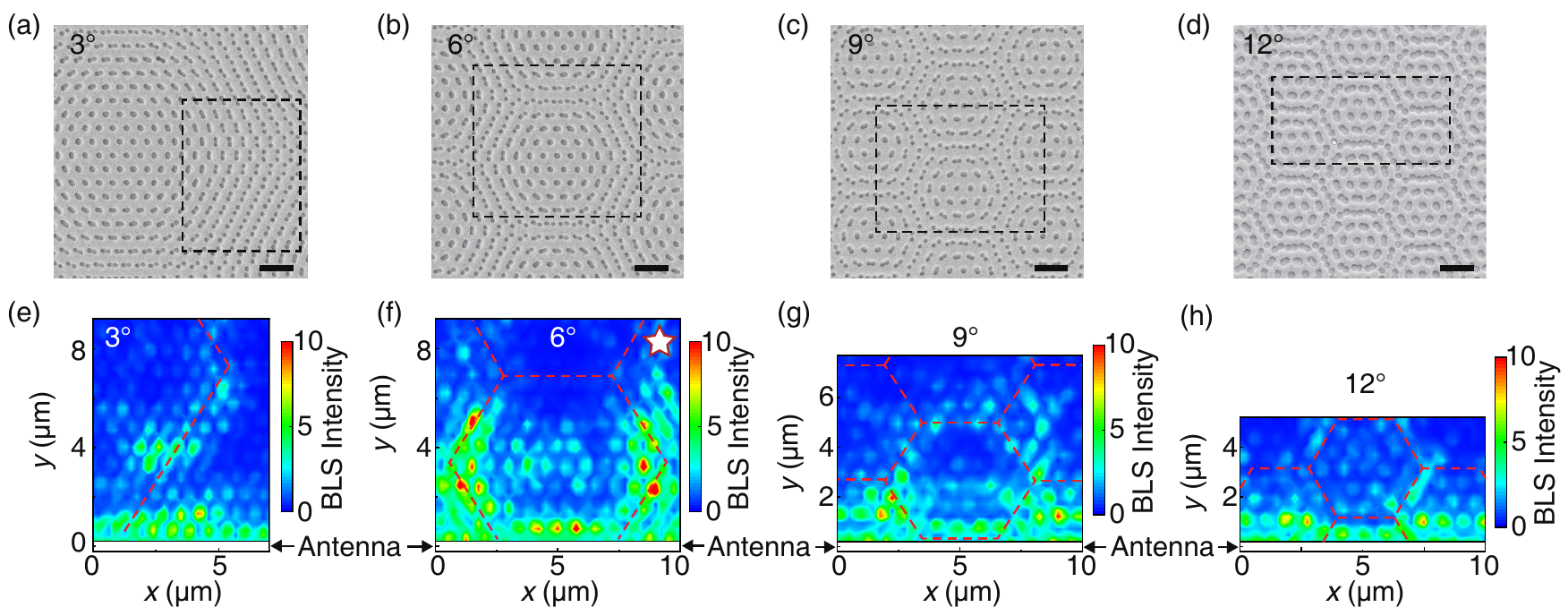}
\caption{(a)-(d) SEM images of moir\'e magnonic lattices with twist angles of 3$^\circ$, 6$^\circ$, 9$^\circ$ and 12$^\circ$, respectively. Black dashed windows are the $\mu$-BLS scanning regions, e.g. 10~$\mu$m\,$\times$\,10~$\mu$m for (b). All scale bars are 2$\,\mu$m. (e-h) Two-dimensional spin-wave intensity maps measured by $\mu$-BLS at a twist angle of 3$^\circ$, 6$^\circ$, 9$^\circ$ and 12$^\circ$ with an excitation frequency of 3.10~GHz and applied magnetic field of 50~mT. Red dashed lines are guide to the eyes for moir\'e unit cells. The white star marks the edge mode profile around the optimal twist angle or `magic angle'.}
\label{fig2}
\end{figure*}

\section{II. Sample information and spin-wave measurements}

The moir\'e magnonic lattices were formed using two sets of antidot triangle lattices meshed into one single YIG layer with different values of the twist angle $\theta$ as shown in Fig.~\ref{fig1}(a). One single antidot lattice acts as a conventional magnonic lattice with a lattice constant $a$\,=\,800\,nm and an antidot diameter of 260~nm. The moir\'e magnonic lattices were fabricated using ebeam lithography and ion beam etching based on an 80$\,$nm-thick YIG film grown by magnetron sputtering at room temperature on 0.5 mm-thick (111) gadolinium gallium garnet (GGG) substrates and annealed at 800$^\circ$C for 4 hours in 1.12 Torr oxygen, subsequently. A gold stripline antenna with a width of 400~nm was then integrated on the moir\'e magnonic lattices to excite spin waves with a microwave source generator. The nanostripline (NSL) provides a broadband excitation~\cite{Ciu2016,vanderSar2020} in wavevector space that covers the first Brillouin zone (BZ) boundary of the magnonic crystal as shown in the antenna fast Fourier transformation (FFT) in the Supplementary Material (SM) Fig.~S1~\cite{SI}. Spin-wave propagation was probed by $\mu$-BLS [Fig.~\ref{fig1}(b)] with a spatial resolution of approximately 250~nm~\cite{Madami_book} (see Appendix A). The $\mu$-BLS signals were measured as a function of excitation frequency in the AB (incommensurate) and AA (commensurate) regions as shown in Fig.~\ref{fig1}(c) with an applied field of 50~mT parallel to the antenna. Apart from the uniform mode around 2.7 GHz, associated with the spatially uniform mode excited through the lattice, another intense peak is detected at 3.1~GHz propagating along the edges of a moir\'e unit cell as shown in Fig.~\ref{fig1}(e) with a contour profile. The raw data of Fig.~\ref{fig1}(e) with a grid profile are presented in the SM Fig.~S2~\cite{SI} for comparison. The moir\'e edge mode is highly sensitive to its excitation frequency (see SM Section~III~\cite{SI}). At the black circle in Fig.~\ref{fig1}(f), an additional mode is observed around 2.9~GHz which is identified as a moir\'e cavity mode. The spatial mapping of a full moir\'e unit cell is presented in SM Fig.~S4~\cite{SI}. Both moir\'e edge and cavity modes detected by the $\mu$-BLS have clear field dependence [Fig.~\ref{fig1}(d)], that agrees well with modes detected by the all-electrical spin-wave spectroscopy based on a vector network analyzer (see Appendix B). 

Figures~\ref{fig2}(a)-(d) present the SEM images of moir\'e magnonic lattices with different twist angles $\theta$ of 3$^\circ$, 6$^\circ$, 9$^\circ$ and 12$^\circ$, for which the moir\'e lattice constants $a_\text{m}$ are calculated to be 15.3~$\mu$m, 7.6~$\mu$m, 5.1~$\mu$m and 3.8~$\mu$m with a single-layer periodicity $a$\,=\,800\,nm, based on the simple estimation $a_\text{m}$=a/$\theta$ ($\theta$ in units of $rad$)~\cite{Dong2021}. NSL antennas for microwave excitation are placed at the bottom of the black dashed windows [Figs.~\ref{fig2}(a-d)] within which the $\mu$-BLS mappings are measured and shown in Figs.~\ref{fig2}(e)-(h). Spin-wave moir\'e edge mode profiles are found to depend critically on two key parameters, i.e. the twist angle $\theta$ and the applied magnetic field $H$ (see Table 1 in the SM Section~V~\cite{SI} for the full diagram). At a certain magnetic field of 50~mT, moir\'e edge mode profiles are measured for different twist angles of 3$^\circ$, 6$^\circ$, 9$^\circ$ and 12$^\circ$ as shown in Figs.~\ref{fig2}(e)-(h) with the same excitation frequency of 3.10~GHz. The edge mode profile optimizes around 6$^\circ$ in terms of both $\mu$-BLS signal intensity and peak linewidth as shown in the SM Fig.~S5~\cite{SI}. This indicates that the twist angle of 6$^\circ$ can be considered as a `magic angle' of the magnonic moir\'e lattice for a given magnetic field of 50~mT in analogy with those in electronic~\cite{Bistritzer2011,Cao2018} and photonic~\cite{Dong2021,Mao2021} moir\'e systems. However, the `magic angle' is not fixed at 6$^\circ$ but sensitive to the external magnetic field, e.g. at 40~mT, the magic angle is 3$^\circ$ while at 60~mT it becomes 9$^\circ$ as indicated by stars in Table 1 of the SM Section~V~\cite{SI}. The applied field may tune the local  magnetization alignment at the ``top" and ``bottom" layers of the moir\'e superlattice that is known to affect the interlayer magnon-magnon coupling strength~\cite{Klingler2018,Chen2018,Qin2018SR,Shiota2020}. Just as the magic angle depends on the interaction strength in electronic moir\'e systems~\cite{Pan2020} and on the interlayer separation in photonic moir\'e crystals~\cite{Wang2020}, the magic angle in magnonic moir\'e lattices depends on the interlattice coupling strength that can be tuned by an applied magnetic field. 

\begin{figure*}
\centering
\includegraphics[width=178mm]{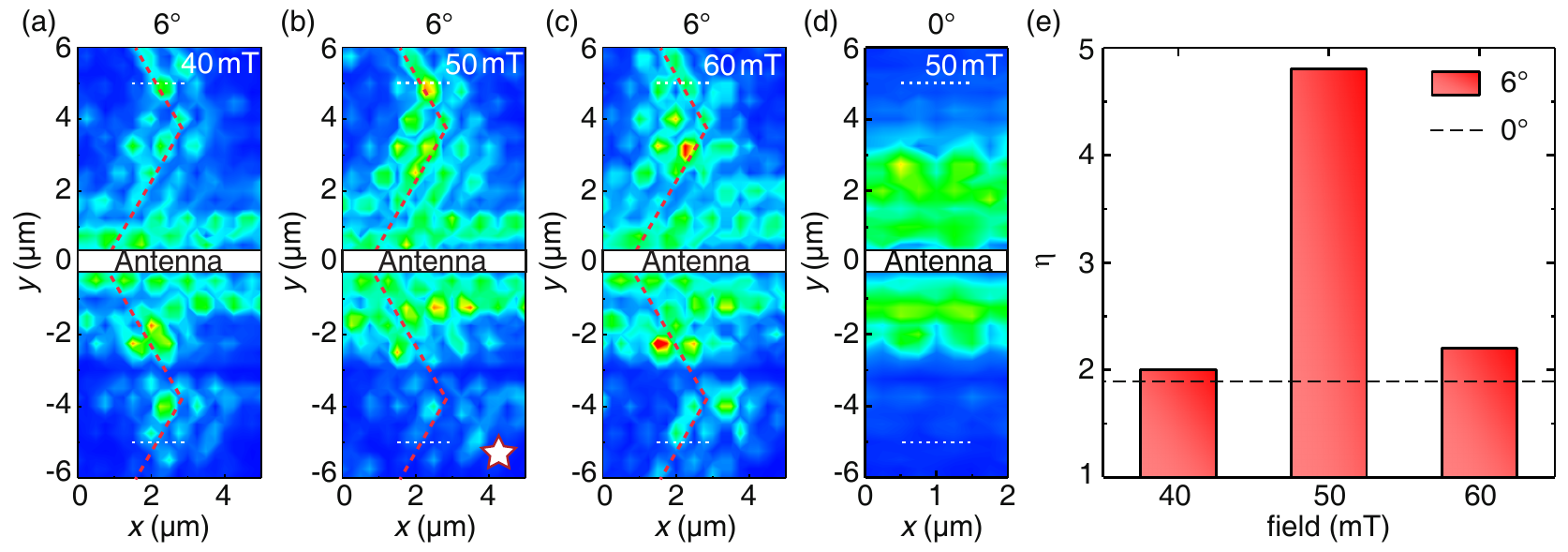}
\caption{Spin-wave edge mode profiles measured by $\mu$-BLS on a magnetic moir\'e lattice with a twist angle of 6$^\circ$ with an external magnetic field of 40~mT (a), 50~mT (b) and 60~mT (c) applied parallel to the stripline antenna (white bar). The excitation frequency is set at 3.1~GHz. The white star denotes the chiral edge mode profile at an optimal applied field value of 50~mT at a fixed twist angle of 6$^\circ$. (d) Spin-wave spatial profile measured by $\mu$-BLS on a conventional magnonic crystal without moir\'e pattern, i.e. zero twist angle with an applied field of 50~mT. The antennas used in (a-d) are of the same design, i.e. 400~nm-wide NSL. (e) Chirality ratio $\eta=I_{+k}/I_{-k}$ extracted from the experimental data for 6$^\circ$ (red columns) and 0$^\circ$ (black dashed line) at a distance of 5~$\mu$m from the antenna [white dotted lines in (a-d)].}
\label{fig3}
\end{figure*}

On the other hand, at a certain twist angle for instance of 6$^\circ$, one can find an optimal applied field (`magic field', if one will) for generating the most intense edge mode profiles as shown in Figs.~\ref{fig3}(a)-(c). Here, another salient feature of the moir\'e edge mode is observed as its chirality in the sense that spin waves propagating towards two opposite directions ($+k$ and $-k$) exhibit different intensities. To evaluate the strength of this effect, we introduce a chirality ratio as $\eta=I_{+k}/I_{-k}$ where $I_{+k}$ and $I_{-k}$ are spin-wave intensities measured by $\mu$-BLS with positive and negative wavevectors $+k$ and $-k$, respectively. At a fixed twist angle of 6$^\circ$, spin-wave edge modes measured at different applied fields (40~mT, 50~mT and 60~mT) show different chiral propagation behavior [Figs.~\ref{fig3}(a-c)], from which the chirality ratios $\eta$ are extracted and presented as the red columns in Fig.~\ref{fig3}(e). At the optimal field (or `magic field') of 50~mT, the chirality ratio maximizes at a value of 4.8 (see SM Fig.~S6~\cite{SI} for the raw data), whereas smaller chirality ratios of 2.0 and 2.2 are found for 40~mT and 60~mT, respectively. Meanwhile, it is also known that a stripline antenna can introduce a certain chirality~\cite{Demi2009,TYu2021b,vanderSar2021} when exciting spin waves in a Damon-Eshbach configuration~\cite{DE1961,Yamamoto2019,Mohseni2019}. To assess and distinguish the chirality induced by the antenna and by the moir\'e system, we conduct a control measurement with the same antenna on a non-moir\'e magnonic crystal, i.e. $\theta$\,=\,0$^\circ$ at 50~mT as shown in Fig.~\ref{fig3}(d), where a weak chirality is observed with $\eta\simeq1.8$ [dashed line in Fig.~\ref{fig3}(e)] attributed purely to the antenna excitation~\cite{Demi2009,TYu2021b,vanderSar2021}. The significantly enhanced chirality ratio at the optimal magnetic field and twist angle may arise from the magnonic band modification by the moir\'e pattern~\cite{Chen2022}.

\begin{figure*}
\centering
\includegraphics[width=178mm]{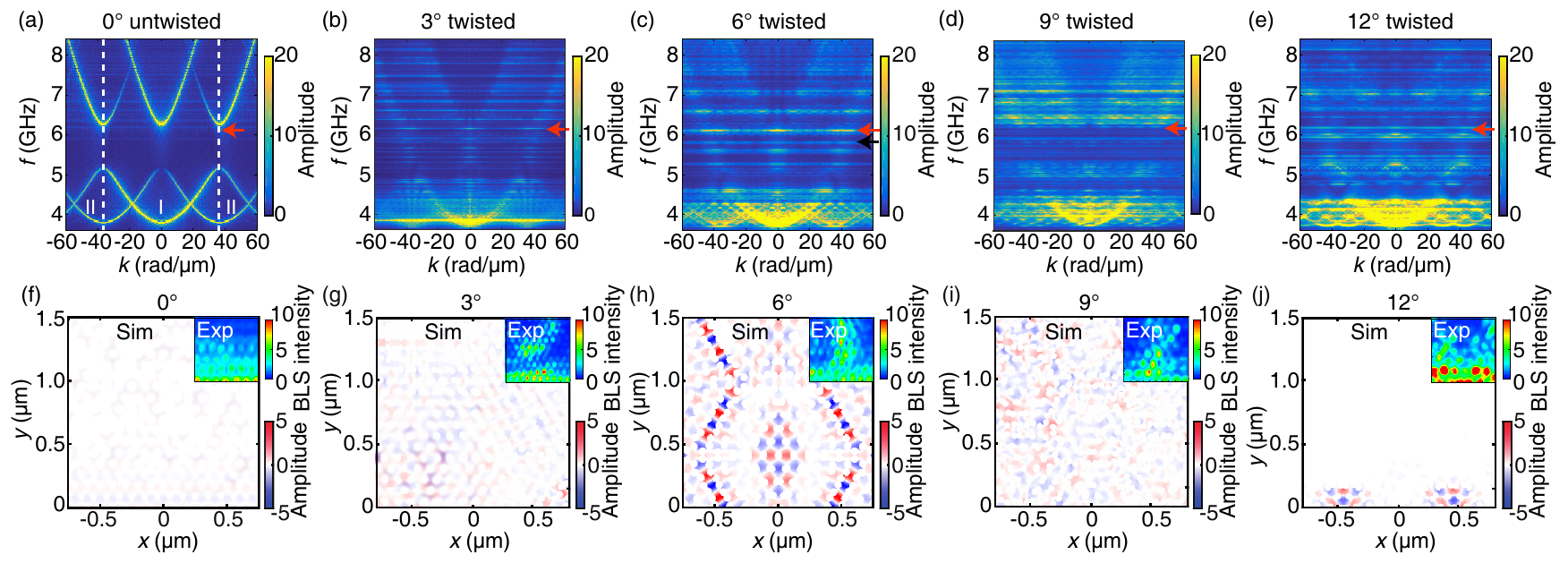}
\caption{(a) Magnonic band structure calculated by micromagnetic simulations for the untwisted antidot magnonic lattice with a periodicity of 100~nm, where the dashed white lines are the BZ boundaries. Spin-wave propagation profile in simulations for the mode slightly below the upper band marked by the red arrow (6.10~GHz) is presented in (f). The inset shows the experimental results taken by $\mu$-BLS on the untwisted magnonic lattices for comparison. (b-e) Magnonic band structure from simulation for the moir\'e magnonic lattice with a twist angle of 3$^\circ$, 6$^\circ$, 9$^\circ$, 12$^\circ$, where the red arrow denotes the edge mode at 6.10~GHz whose propagation profile is shown in (g-h), respectively. The insets show the edge mode profile measured by $\mu$-BLS on a sample with a lattice contant of 800~nm. The black arrow in (c) denotes the cavity mode at 5.90~GHz. The magnetic fields in these simulations are set as 50~mT.}
\label{fig4}
\end{figure*}

\section{III. Micromagnetic simulations and theoretical analysis}

\begin{figure*}
\centering
\includegraphics[width=140mm]{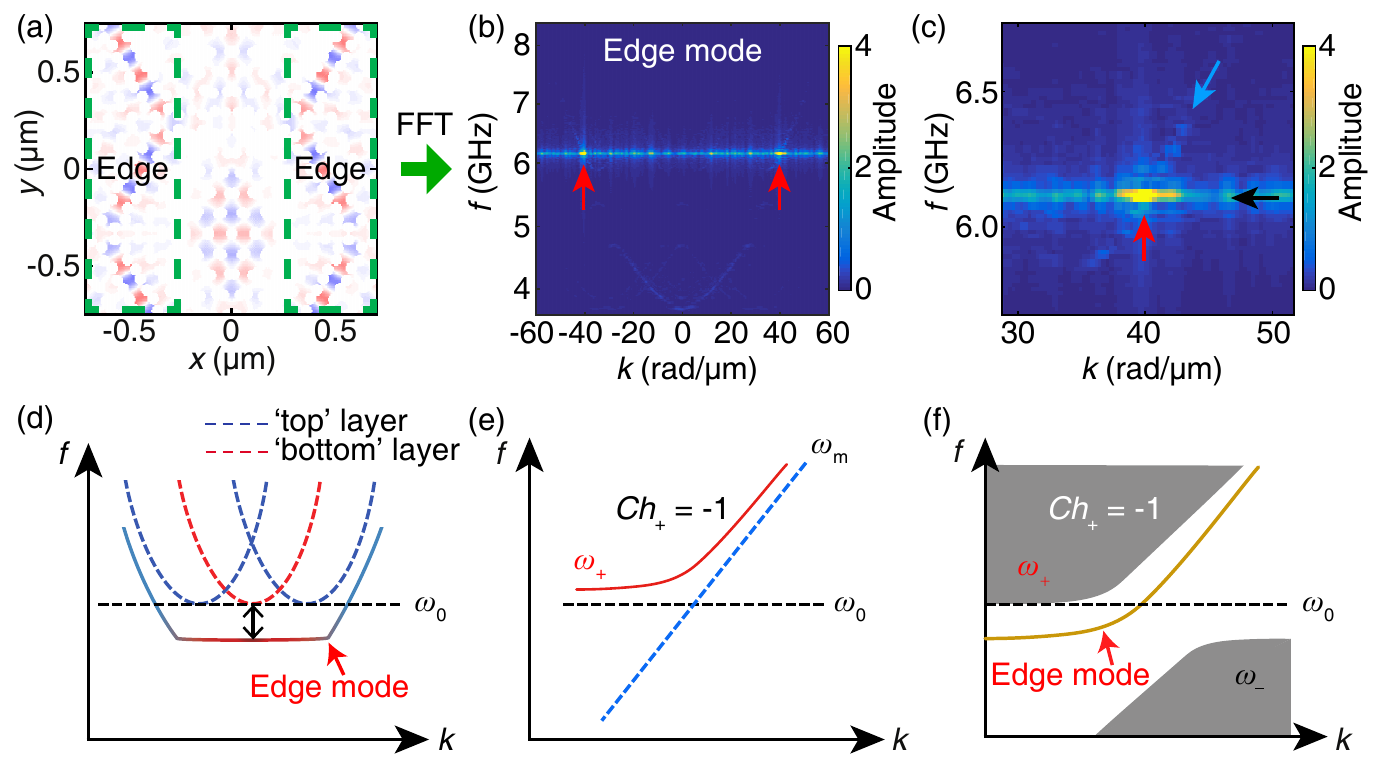}
\caption{(a) Simulated spatial spin-wave intensity mappings on the moir\'e magnonic lattices with a twist angle of 6$^\circ$ at the edge-mode frequency of 6.10~GHz. The edge mode regions within the green dashed squares are taken for the FFT to generate the intensity distribution of the spin-wave dispersion in (b). The red arrows denote the pronounced excitation of the edge modes. (c) The zoom-in plot indicates that the edge mode (red arrow) appears at the intersection of the flatband (black arrow) and a propagating mode (light blue arrow). (d) Sketches for the formation of the mini-flatband and edge mode resulted from mode hybridization of upper bands of two magnonic crystals twisted with respect to one another. The edge mode emerges slightly below the bottom of the upper bands ($\omega_0$) as indicated by the red arrow. (e) Simplified scenario for magnon-magnon coupling between the upper band bottom ($\omega_{0}$) and propagating mode ($\omega_{\rm m}$) considered in the theoretical analysis. Chern number ($Ch_{+}$) for the hybridized upper branch $\omega_{+}$ is calculated to be $-1$. (f) Schematics of the emergent chiral edge mode in the band gap between $\omega_{+}$ and $\omega_{-}$.}
\label{fig5}
\end{figure*}

To further understand the spin-wave edge modes observed by $\mu$-BLS spectroscopy, we perform micromagnetic simulations for structures with twist angles of 0$^\circ$, 3$^\circ$, 6$^\circ$, 9$^\circ$ and 12$^\circ$ based on OOMMF~\cite{oommf}. In order to limit the computing time, we considered antidot magnonic lattices with a periodicity of 100~nm instead of 800~nm for real samples. It is worth noting that the magnon band structure and edge mode spatial profile simulated for 800~nm is primarily consistent with those for 100~nm but demands a significantly longer time for computing even with lower wavevector resolution (see SM Section~VIII~\cite{SI}). For simplicity, the simulation is set in an area (4~$\mu$m$\times$12~$\mu$m) of a 100 nm-period antidot triangle lattice with a hole diameter of 50~nm. The external magnetic field of 50~mT is set along $x$ direction (see Appendix C for more details and parameters). The magnonic band structure is then obtained from simulations first for the untwisted magnonic lattice with $\theta$\,=\,0$^\circ$ [Fig.~\ref{fig4}(a)], where magnonic band gaps~\cite{Neusser2010,Tacchi2011,Barman2021} are clearly observed between 5.15~GHz and 6.25~GHz. The spin-wave mode at 6.10~GHz (red arrows) locates within the forbidden band gap (see Fig.~\ref{fig4}(a) and its zoom-in spectra in the SM Fig.~S8~\cite{SI}) and therefore can hardly be excited as shown in Fig~\ref{fig4}(f). With a twist angle of 6$^\circ$ [Fig.~\ref{fig4}(c)], the magnonic band structure is completely modified after complex mode hybridization. Instead of a clear band gap, several mini-flatbands~\cite{Chen2022} [red and black arrows in Fig.~\ref{fig4}(c)] arise near the first BZ boundaries due to the magnon-magnon hybridization~\cite{Chen2018,Klingler2018,Qin2018SR,Shiota2020}, as illustrated in Fig.~\ref{fig5}(d). The moir\'e edge mode emerges at 6.10~GHz (red arrow) with its spatial profile shown in Fig.~\ref{fig4}(h), whereas the moir\'e cavity mode appears at 5.80~GHz (black arrow) with its spatial profile shown in the SM Fig.~S9(a)~\cite{SI}. However, when the twist angle is tuned up to 12$^\circ$, the mini-flatband starts to disappear by losing its flatness [Fig.~\ref{fig4}(e)], and consequently, the quality of moir\'e edge mode deteriorates severely [Fig.~\ref{fig4}(j)]. The cavity mode around 5.80~GHz [black arrows in Figs.~\ref{fig4}(c)] is well-confined at the AA region for $\theta$\,=\,6$^\circ$, but becomes more scattered for $\theta$\,=\,12$^\circ$ as shown in the SM~Fig.~S9(b)~\cite{SI}. The formation of the mini-flatband appears as a result of mode hybridization induced by the interlayer magnon-magnon coupling~\cite{Chen2018,Klingler2018,Qin2018SR,Shiota2020} between two magnonic-crystal layers~\cite{Chen2022} in analogy to flat bands formed by interlayer interaction in its electronic~\cite{Bal2020} and photonic~\cite{Dong2021} counterparts.

To investigate the origin of the edge mode, we take the real-space simulation results of Fig.~\ref{fig4}(h) and perform fast Fourier transformation (FFT) at the AB regions in which edge modes locate as shown in Fig.~\ref{fig5}(a). The magnon edge modes correspond to two strong intensities in reciprocal space at the intersection between the flatband and an ordinary propagating mode as indicated by the red arrows in Fig.~\ref{fig5}(b) and shown in a blow-up dispersion in Fig.~\ref{fig5}(c). The actual formation process of edge mode and flatbands may be rather complicated and requires more sophisticated theoretical investigation in the future. Here we consider a simplified scenario for the edge mode formation as illustrated in Fig.~\ref{fig5}(d), where the magnonic upper band of the `bottom' layer (red dashed line) hybridizes with that of the `top' layer (blue dashed lines) twisted with respect to the `bottom' one and form a new magnon band structure with edge modes located at the turning point. To enable further theoretical calculation, we assume that the hybridization process occurs in two steps: 1) The minimums of the upper bands of both layers (bottom of red and blue dashed lines) first form a ``flatband" $\omega_0$~\cite{Chen2022}. 2) The flatband $\omega_0$ couples subsequently with the propagating mode $\omega_{\text{m}}$. The second step is illustrated in Fig.~\ref{fig5}(e). The simulation results in Fig.~\ref{fig5}(c) suggest that the edge mode corresponds to the pronounced crossing point between a flatband mode and a propagating mode with a positive group velocity. Based on this observation, we consider the edge mode result from the mode hybridization shown in Figs.~\ref{fig5}(e) and (f). In the following, we demonstrate theoretically that the upper branch $\omega_{+}$ formed by hybridization exhibits a nontrivial topology with a Chern number $Ch_{+}=-1$ and the edge mode appears in the band gap formed after the hybridization.

We adopt the theoretical approach used in Ref.~\cite{Okamoto2020,Okamoto20202} to study the Berry curvature induced by the magnon-photon hybridization and apply it to the magnon-magnon coupling scenario of this work. Here, an initial approximation is necessary where we separate the merged moir\'e magnonic superlattice into two adjacent antidot sublattices twisted with one another (see its illustration in SM~Fig.~S10~\cite{SI}), so that the coupling between two twisted lattices can be considered as \textit{interlayer} magnon-magnon coupling mediated by the dipolar interaction~\cite{Chen2018,Shiota2020}. A simplified model is considered where only the coupling between the neighbouring antidots with the same registry in each layer is calculated. Each antidot can be regarded as a macroscopic magnetic dipole under an in-plane field suggested by the simulation shown in the SM Section~XII~\cite{SI}. After a lengthy calculation (see Appendix D), we obtain the Berry curvature $\Omega_{z,+}\left(k\right)$ for the hybridized mode $\omega_{\rm +}$ [Fig.~\ref{fig5}(e)] as, 
\begin{equation}\label{BerryCurve}
\Omega_{z,+}(k)=\frac{1}{k}\frac{\partial}{\partial k}\left[\frac{\epsilon_{\rm d}^2(\epsilon_+^3+\epsilon_+)}{2\epsilon_{\rm d}^2(2\epsilon_+^2-1)+2\epsilon_{\rm m}(\epsilon_+^2-1)^2}\right],
\end{equation}
where $\epsilon_+=\omega_+/\omega_{0}$, $\epsilon_{\rm m}=\omega_{\rm m}/\omega_{0}$, and $\epsilon_{\rm d}=\omega_{\rm d}/\omega_{0}$ with $\omega_{\rm d}$ represents the term associated with the dipolar interaction (see Appendix D). The Chern number $Ch_+$ for the edge mode at the hybridized point of $\omega_+$ can then be calculated by the integral of the Berry curvature [Eq.~\ref{BerryCurve}] over the two-dimensional wavevector space~\cite{Thouless1982} using,
\begin{equation}
\label{Chern}
Ch_+=\frac{1}{2\pi}\int_{\rm BZ}\Omega_{z,+}(k)dk_xdk_y.
\end{equation}
The Chern number $Ch_{\rm +}$ for the hybridized magnon mode $\omega_{\rm +}$ can be derived based on Eq.~\ref{Chern} to be $Ch_{\rm +}=-1$ (see Appendix D). This reveals the non-trivial topological nature~\cite{Shindou2013} of the magnon edge mode in moir\'e magnonic lattices. Through the calculation process, the external magnetic field can tune the local magnetization orientation and affect the interlattice dipolar interaction that may lead to a change on the Berry curvature of the hybridized system. As a result, the chiral edge mode is responsive to the applied magnetic field as observed in experiments [Figs.~\ref{fig3}(a-c)], which resonates with a recent theoretical study showing tunable magnonic Chern bands with an external magnetic field~\cite{Luqiao2022} based on dipolar interaction in multilayers. Considering the fact that the mini-flatband $\omega_{0}$ depends critically on the twist angle $\theta$, the non-trivial topological spin-wave edge modes emerge at a certain combination of the twist angle $\theta$ and magnetic field $H$, which accounts for the spin-wave edge modes observed by $\mu$-BLS in experiments (see Table 1 in the SM Section~V~\cite{SI}). The microscopic mechanism of the nontrivial topological spin-wave edge mode is the dipolar interaction~\cite{Shindou2013,Li2018} which relies on the relative position of two spins (tunable by twist angle $\theta$) and their spin orientations (tunable by magnetic field $H$)~\cite{Shen2020}. Our conclusion that dipolar interaction is responsible for our observation resonates with the origin of topological spin-wave edge modes theoretically predicted in other magnonic lattices~\cite{Shindou2013,Li2018}. 

\section{IV. Discussion and Conclusion}

The observed topological spin-wave edge modes emerge at the edge of a moir\'e unit cell that is analogous to the topological mosaics predicted in twisted van der Waals bilayers~\cite{Tong2017}. Remarkably, if one rotates the wavevector excitation towards the $M$ point instead of the $K$ point symmetry in the reciprocal space, spin waves tend to propagate along the edge of the entire lattice referred to as `bulk edge mode' in the simulation results of Fig.~S12 in the SM~\cite{SI}. In this work, however, we focus on investigating the moir\'e edge mode [Fig.~\ref{fig1}(e)] rather than the bulk edge mode which we leave for future experimental and theoretical investigations. Further simulation results reveal that the quality of the edge mode relies also on the diameter of the antidots as shown in the SM Fig.~S13~\cite{SI}. This indicates that the filling factor is an additional parameter for the generation of non-trivial topological magnon modes comparable with the important role of filling ratio in a topological phononic system as demonstrated recently in a numerical study~\cite{Pirie2022}. Although we attribute the topological origin of our observation to be primarily interlattice dipolar interaction, the interlayer exchange (RKKY) interaction~\cite{Parkin1990} and Dyzaloshinskii-Moriya interaction~\cite{Gam2021} in magnetic moir\'e bilayer systems may also play a role in forming spin-wave edge modes and is challenging while interesting to investigate in the future for example theoretically using tight-binding models~\cite{Shen2018,Eck2022}.

In conclusion, spin-wave edge modes and cavity modes are experimentally demonstrated in moir\'e magnonic lattices by micro-focused BLS. The edge modes are observed at the boundary of a moir\'e unit cell while the cavity modes are localized within the center of the moir\'e unit cell. With an applied field of 50~mT, the moir\'e edge mode is most intense at a magic twist angle of 6$^\circ$. The dependence of the `magic angle' on the applied magnetic field indicates that the dipolar interaction between the twisted magnonic sublattices plays an important role in the formation of the moir\'e edge modes. The magnetic field offers an additional degree of freedom for tuning the magnon edge mode on top of the twist angle and thus provides more versatility to magnonic moir\'e devices. The micromagnetic simulations show that the edge mode arises at the crossing point between a moir\'e flatband and a propagating magnon branch near the first BZ boundary. Estimates of the Berry curvature for magnon-magnon coupling further confirms that the dipolar interaction is the key mechanism of magnon edge modes that exhibit a non-trivial topological nature with a non-zero integral Chern number. The moir\'e spin-wave edge modes observed in this work, as the magnonic counterpart of the magic-angle electronic and photonic systems, open an emergent research direction of moir\'e magnonics. The use of topologically protected magnon edge modes will greatly expand the functionalities of magnonic devices for information processing. 

\section{Acknowledgments}
We wish to acknowledge the support by the National Key Research and Development Program of China Grant No.~2022YFA1402801, by NSF China under Grant Nos. 12074026, 52225106 and U1801661, by China Scholarship Council (CSC) under Grant No.~202206020091 and by Shenzhen Institute for Quantum Science and Engineering, Southern University of Science and Technology (Grant No. SIQSE202007). M.M. and G.G. acknowledge financial support from the Italian Ministry of University and Research through the PRIN-2020 project entitled `The Italian factory of micromagnetic modelling and spintronics', cod. 2020LWPKH7.

\section{Appendix A: Micro-focused Brillouin light scattering}
Brillouin light scattering spectroscopy is based on the inelastic scattering of light by thermally or $rf$ excited spin waves. A diffraction limited spatial resolution of about 250 nm is reached by employing a single-mode solid-state laser with a wavelength of 532 nm, at normal incidence, using a microscope objective with numerical aperture 0.75. A (3+3)-pass tandem Fabry-P\'erot interferometer is used to analyze the inelastically scattered light. A nanopositioning stage allows to position the sample with a precision down to 10 nm on all three axes. A DC/AC electrical probe station ranging from DC up to 20 GHz is used for spin-wave pumping. The microwave power is set at +18 dBm on the rf generator output.

\section{Appendix B: All-electrical spin wave spectroscopy}
The all-electric spin-wave spectroscopy consists of Rohde $\&$ Schwarz VNA with frequency range from 10 MHz to 40 GHz and a U-shaped electric-controlled external magnets. The YIG film is put between two magnetic poles. The external magnetic field is parallel with the NSL antenna and perpendicular to the spin-wave wavevectors. The integrated electrodes of the antenna on the top of moir\'e magnonic crystals are connected to the VNA via the microwave probes and cables. Therefore, the microwave currents can be injected into the excitation antennas and we measure the reflection spin-wave spectra of the nanostructured moir\'e magnonic crystals with sweeping the external magnetic field from -150 mT to 150 mT, after first saturating the film at -300 mT. The IF bandwidth is set to be 1 kHz.

\section{Appendix C: Micromagnetic simulations}
The OOMMF (http://math.nist.gov/oommf) program is used for the micromagnetic simulations. Two-dimensional periodical boundary condition is considered in the simulation. The size of the single mesh cell is 5 nm$\times$ 5 nm. Two twisted triangle magnonic crystals are meshed into one layer to form the moir\'e pattern. We set the saturation magnetization $M_{\rm S}$= 140 kA/m, damping $\alpha$=10$^{-4}$ and the exchange coefficient $A$=3.7$\times$10$^{-12}$ J/m for YIG films. A 20 nm-wide stripline antenna is considered with a dynamic field in $x$ direction, which is described by ($H_{\rm ex}=h_0\sin(2\pi f(t-t_0))/(2\pi f(t-t_0))$), where the excitation field $h_0$=2 mT, $f$=20 GHz and $t_0$=100.1 ps. The dynamic field contains same excitation strength from 0 to $f$ GHz. The external magnetic field of 50 mT is set along $x$ direction, which is perpendicular to the wavevector direction along $y$. The magnetization ground state is calculated by minimizing the total energy of the YIG moir\'e magnonic crystals, based on which the magnetization dynamics is simulated for 1050 equidistant times with a time step of 25 ps. We then perform the two-dimensional fast Fourier transformation along $y$ direction to obtain the magnon band structures. For single-frequency excitation, we use $H_{\rm ex}=h_0\sin(2\pi f_{\rm fixed}(t-t_0))$, where $f_{\rm fixed}$ is the fixed excitation frequency. The spatial maps present the $x$ component of the magnetization dynamics at a certain time point.

\section{Appendix D: Theoretical calculations}
Taking into account the dipolar interaction in the magnonic moir\'e lattices, we utilize Landau-Lifshitz equations to analyze the magnetization dynamics as
\begin{equation} \label{llg}
	\begin{aligned}
		&\frac{d \bm{m}}{d t}=-\gamma \mu_{0} \bm{m} \times\left(\bm{H}_{{0}}+\bm{H}_{{\rm fb}}+\bm{H}_{{\rm dip}}\right), \\
		&\frac{d \tilde{\bm{m}}}{d t}=-\gamma \mu_{0} \tilde{\bm{m}} \times\left(\bm{H}_{{0}}+J_{0} \nabla^{2} \tilde{\bm{m}}+\tilde{\bm{H}}_{{\rm dip}}\right).
	\end{aligned}
\end{equation}
Here, $\bm{H}_{0}$ is the external magnetic field, $\bm{H}_{\rm fb}$ is the phenomenological field which creates a moir\'e flat band, $J_{0}$ is the exchange stiffness. In this appendix, the lattice-mode parameters are marked with a tilde.	Since we only concern about the eigen-frequencies, we disregard the damping terms.

Then we estimate the hybridization of the lattice mode and mini-flat band. The Hamiltonian of interlayer dipole-dipole interaction under our approximation reads
\begin{equation}
	\mathcal{H}_{\rm dip} = \frac{\mu_0 }{2} \sum_{i\neq j} \frac{R_{ij}^2 \bm{m}_i \cdot \bm{m}_j-3\left(\bm{R}_{ij} \cdot \bm{m}_i \right) \cdot \left(\bm{R}_{ij} \cdot \bm{m}_j \right) }{R_{ij}^5},
\end{equation}
where $\bm{R}_{ij}$ represents the dislocations of antidots, and for nearest neighbors, we define $R_{ij} = R$. Dipolar fields $\bm{h}_{\rm dip}$ and $\tilde{\bm{h}}_{\rm dip}$ arises from this Hamiltonian. The total dipolar field of a macroscopic dipole is represented by $\bm{H}_{\rm dip}$, whose scale is accumulated by the antidot's volume. As mentioned before, we only consider the nearest neighbor's contribution. The dipolar effective field $\bm{H}_{\rm dip}$ can be estimated by
$
\bm{H}_{\rm dip} =\frac{1}{4}\bm{h}_{\rm dip}\cdot \pi \Phi^{2} t,
$
where $t$ is the thickness of the film and $\Phi$ is the diameter of the holes. 
For antidots from two different layers locating at the moir\'e unit cell edge, this parameter $R$ could be written approximately as:
\begin{equation}
	R \cong \frac{\lambda}{2} \theta_{{\rm twist}}=\frac{a}{2 \theta_{{\rm twist}}} \cdot \theta_{{\rm twist}}=\frac{a}{2}
\end{equation}
where the $\theta_{\rm twist}$ is the twisted angle between two layers. Therefore, the static dipolar field is estimated by $H_{\rm dip}=\tilde{H}_{\rm dip}=\frac{M_{\rm S}}{16R^{3}}\Phi^{2}t$. It affects the orientations of antidots' magnetizations, which means that the static orientation of $\bm{m}$ and $\tilde{\bm{m}}$ might be not inherently following the direction of the external magnetic field. To make the total energy minimum under the coexistence of forementioned dipolar interaction, local shape anisotropy, and external field, it comes to an equilibrium static state with spatial distributions. In other words, the external field $\bm{H}_{0}$ enforces magnetization to be fixed, while it is not strong enough, with respect to the static dipolar field, to predominate the direction of saturated magnetization orientations. We define the angle between magnetization orientation and the external field as $\beta$.
Thus, equations of motion (EOM) describing the antidot moir\'e spin waves are written as:
\begin{widetext}
	\begin{equation}
		\left\{\begin{array}{ll}
			\frac{d m_{ x}}{d t}=-\gamma \mu_{ \rm0}\left[H_{ \rm0} \cos \beta+H_{ \rm{fb}}-H_{ \rm{dip}}\left(1-3 \sin ^{2} \phi\right)\right] m_{ y}-\gamma \mu_{ \rm0} H_{ \rm{dip}} \tilde{m}_{ y}, \\
			\frac{d m_{ y}}{d t}=\gamma \mu_{ \rm0}\left[H_{ \rm0} \cos \beta+H_{ \rm{fb}}-H_{ \rm{dip}}\left(1-3 \sin ^{2} \phi\right)\right] m_{ x}+\gamma \mu_{ \rm0} H_{ \rm{dip}}\left(1-3 \cos ^{2} \phi\right) \tilde{m}_{ x}+C, \\
			\frac{d \tilde{m}_{ x}}{d t}=-\gamma \mu_{ \rm0}\left[H_{ \rm0} \cos \beta+J_{ \rm0} M_{ \rm{S}}|k|^{2}-\tilde{H}_{ \rm{dip}}\left(1-3 \sin ^{2} \phi\right)\right] \tilde{m}_{ y}-\gamma \mu_{ \rm0} \tilde{H}_{ \rm{dip}} m_{ y}, \\
			\frac{d \tilde{m}_{ y}}{d t}=\gamma \mu_{ \rm0}\left[H_{ \rm0} \cos \beta+J_{ \rm0} M_{ \rm{S}}|k|^{2}-\tilde{H}_{ \rm{dip}}\left(1-3 \sin ^{2} \phi\right)\right] \tilde{m}_{ x}+\gamma \mu_{ \rm0} \tilde{H}_{ \rm{dip}}\left(1-3 \cos ^{2} \phi\right) m_{ x}+C,
		\end{array}\right.
	\end{equation}
\end{widetext}
where $\phi=\pi/6+\beta$.
In these equations, there are inhomogeneous terms $C$ which shouldvanish since we are looking for the linear response. Thus, we find that the angle $\beta$ obeys the following equation:
\begin{equation}
	\label{inhomo}
	C=-H_{0} M_{{\rm S}} \sin \beta+3 H_{{\rm dip}} M_{{\rm S}} \sin \phi \cos \phi=0.
\end{equation}

Then, we let $\bm{x}_{k}=\left(m_{x},m_{y},\tilde{m}_{x},\tilde{m}_{y}\right)^{\intercal}$ represent the wavefunction of spin waves. Our EOM could be transformed as:
\begin{equation}
	i\frac{d\bm{x}_{k}}{d t} =  \mathcal{H}_{\rm{eff}}\cdot\bm{x}_{k}.
\end{equation}
Then, we define parameters as follows:
\begin{equation}
	\left\{
	\begin{array}{l}
		\omega_{\rm d}=\gamma \mu_{ \rm 0} H_{ \rm dip}, \\
		\omega_{ \rm 0}=\gamma \mu_{ \rm 0}\left(H_{ \rm 0}+H_{ \rm fb}\right)-\omega_{ \rm d}\left(1-3\sin^2{\phi}\right),\\
		\omega_{ \rm m} = \gamma \mu_{ \rm 0}\left(H_{ \rm 0}+J_{ \rm 0} M_{ \rm S}\tilde{k}^2\right)-\omega_{ \rm d}\left(1-3\sin^2{\phi}\right),
	\end{array}
	\right.
\end{equation}
and under these definitions, our effective Hamiltonian reads
\begin{widetext}
	\begin{equation}
		\mathcal{H}_{\rm{eff}}=\left(\begin{array}{cccc}
			0 & -i \omega_{{\rm 0}} & 0 & -i \omega_{{\rm d}} \\
			i \omega_{{\rm 0}} & 0 & i \omega_{{\rm d}}\left(1-3 \cos ^{2} \phi\right) & 0 \\
			0 & -i \omega_{{\rm d}} & 0 & -i {\omega}_{{\rm m}} \\
			i \omega_{{\rm d}}\left(1-3 \cos ^{2} \phi\right) & 0 & i {\omega}_{{\rm m}} & 0
		\end{array}\right).
	\end{equation}
\end{widetext}
Nonetheless, this Hamiltonian could be spotted easily as a non-Hermitian one. Within a energy conservation system, the Hamiltonian should be Hermitian. This information implies a non-zero Berry curvature. Then we assume a Hermitian matrix $\sigma$ that,
\begin{equation}
	\sigma=\left(\begin{array}{c c c c}
		0 & -i & 0 & 0 \\
		i & 0 & 0 & 0 \\
		0 & 0 & 0 & -i \\
		0 & 0 & i & 0
	\end{array}\right),
\end{equation}
and define $\mathcal{\tilde {H}}_{\rm{eff}}=\sigma \cdot  \mathcal{H}_{\rm{eff}}$. Then $\mathcal{\tilde {H}}_{\rm{eff}}$ is Hermitian. Actually, it is real and symmetric, with, 
\begin{equation}\label{fixedH}
	\mathcal{\tilde {H}}_{\rm{eff}}=\left(\begin{array}{cccc}
		\omega_{ \rm {0}} & 0 & \omega_{ \rm {d}}\left(1-3 \cos ^{2} \phi\right) & 0 \\
		0 & \omega_{ \rm {0}} & 0 & \omega_{ \rm {d}} \\
		\omega_{ \rm {d}}\left(1-3 \cos ^{2} \phi\right) & 0 & \omega_{ \rm {m}} & 0 \\
		0 & \omega_{ \rm {d}} & 0 & \omega_{ \rm {m}}
	\end{array}\right) .
\end{equation}

Now we want to calculate the topological parameters of hybridized bands. According to our experimental results, there is an exceptional point for external magnetic field that could generate edge modes. We intuitively assume that when $1-3\cos^{2}{\phi} = 0$ the conditions might be exceptional. This assumption is not entirely precise but could reveal some nature about the internal mechanism of the field-dependent phase transition. 

The external magnetic field modulate the spin orientations of antidots. From Eq.~\ref{inhomo}, we can analytically find out that the angle $\phi$'s relation to the external field strength $\bm{H}_0$. We note that $\phi=\phi(H_{0})$. 

At this stage, the effective Hamiltonian in Eq.~\ref{fixedH} is now, 
\begin{equation}
	\mathcal{H}_{\rm{eff}}^{\star}=\left(\begin{array}{cccc}
		0 & -i \omega_{{\rm 0}} & 0 & -i \omega_{{\rm d}} \\
		i \omega_{{\rm 0}} & 0 & 0 & 0 \\
		0 & -i \omega_{{\rm d}} & 0 & -i {\omega}_{{\rm m}} \\
		0 & 0 & i {\omega}_{{\rm m}} & 0
	\end{array}\right).
\end{equation}
The eigenvalue and eigenvector equations of the fixed effective Hamiltonian can be written as:
\begin{equation}  \label{eigeneq}
	\sigma \cdot \omega \bm{x}_{{k}}=\mathcal{\tilde {H}}_{\rm{eff}}^{\star} \cdot \bm{x}_{{k}}.
\end{equation}
By solving this equation, the eigenvalues of the fixed effective Hamiltonian are:
\begin{equation}
	\omega_{\pm}^{2}=\frac{\omega_{{\rm 0}}^{2}+\omega_{{\rm m}}^{2}}{2} \pm \sqrt{\left(\frac{\omega_{{\rm 0}}^{2}-\omega_{{\rm m}}^{2}}{2}\right)^{2}+C_{{\rm d}}^{2}},
\end{equation}
where $C_{\rm d}$ represents the coupling strength induced by the dipolar interaction,
\begin{equation}
	C_{\rm d}^{2}=4\omega_{\rm d}^{2}\omega_{\rm 0}\omega_{\rm m}\sim g^{4}.  
\end{equation}
Substituting $\omega_+$ for the eigenvalue $\omega$ in the eigenvalue equation \ref{eigeneq}, we get the eigenfunction $\bm{x}_k$:
\begin{equation}
	\bm{x}_{ k}=\left(\begin{array}{c}
		-i \omega_{ \rm {d}} \omega_{ \rm +}^{2} \\
		\omega_{ \rm {d}} \omega_{ \rm {0}} \omega_{ \rm +} \\
		-i\left[\omega_{ \rm {d}}^{2} \omega_{ \rm {0}}+\left(\omega_{ \rm +}^{2}-\omega_{ \rm {0}}^{2}\right) \omega_{ \rm {m}}\right] \\
		\omega_{ \rm +}\left(\omega_{ \rm +}^{2}-\omega_{ \rm {0}}^{2}\right)
	\end{array}\right).
\end{equation}
According to the methodology developed firstly by Refs~\cite{Okamoto2020,Okamoto20202}, Berry curvature of the coupled waves can be represented as 
\begin{equation}
	\Omega_{{z},+}(k)=\frac{1}{k} \frac{\partial}{\partial k}\left(\frac{\tilde{\bm{x}}_{{k}}^{\dagger} \Sigma \tilde{\bm{x}}_{{k}}}{\tilde{\bm{x}}_{{k}}^{\dagger} \sigma \tilde{\bm{x}}_{{k}}}\right),
\end{equation}
where $\Sigma = \rm{diag}\left(\emph I_{2\times2},\emph O_{2\times2}\right)$, with $I$ representing the unit matrix and $O$ representing zero matrix.
Then, we can calculate the Berry curvature that,
\begin{equation}\label{Berry}
	\Omega_{ \rm +}=\frac{1}{k} \frac{\partial}{\partial k}\left(\frac{\epsilon_{ \rm {d}}^{2}\left(\epsilon_{ \rm +}^{3}+\epsilon_{ \rm +}\right)}{2 \epsilon_{ \rm {d}}^{2}\left(2 \epsilon_{ \rm +}^{2}-1\right)+2 \epsilon_{ \rm {m}}\left(\epsilon_{ \rm +}^{2}-1\right)^{2}}\right).
\end{equation}
We define $N(k)=\frac{\epsilon_{ \rm {d}}^{2}\left(\epsilon_{ \rm +}^{3}+\epsilon_{ \rm +}\right)}{2 \epsilon_{ \rm {d}}^{2}\left(2 \epsilon_{ \rm +}^{2}-1\right)+2 \epsilon_{ \rm {m}}\left(\epsilon_{ \rm +}^{2}-1\right)^{2}}$
by the equation~\ref{Berry}. The upper branch $\omega_{+}$, which nearing but above the observed edge mode, is mainly studied in our following calculations. The Chern number can be written as
\begin{equation}
	C h_{+}=\frac{1}{2 \pi} \int_{\rm B Z} \Omega_{+} \cdot k \cdot d k \cdot d \theta \cong N(\infty)-N(0).
\end{equation}
When $k\rightarrow\infty$, we have $\omega_{+}\rightarrow\omega_{\rm m}\rightarrow\infty$, as well as $\omega_{\rm d}$ and $\omega_{\rm fb}$ are still finite constants, so that $\omega_{\rm 0}/\omega_{\rm m}=\omega_{\rm d}/\omega_{\rm m}=0, (k\rightarrow\infty)$. Therefore, we could simplify that
\begin{equation}
	\begin{aligned}
		N(\infty)&=\frac{\epsilon_{ \rm {d}}^{2}\left(\epsilon_{ \rm +}^{3}+\epsilon_{ \rm +}\right)}{2 \epsilon_{ \rm {d}}^{2}\left(2 \epsilon_{ \rm +}^{2}-1\right)+2 \epsilon_{ \rm {m}}\left(\epsilon_{ \rm +}^{2}-1\right)^{2}} \\
		&\sim \frac{\epsilon_{ \rm {m}}^{3} \epsilon_{ \rm {d}}^{2}}{2\left(\epsilon_{ \rm {m}}^{2} \epsilon_{ \rm {d}}^{2}+\epsilon_{ \rm {m}}^{5}\right)} \rightarrow 0.
	\end{aligned}
\end{equation}
For the weak coupling regime as we assumed, when $k \rightarrow 0$, the approximation reads
\begin{equation}
	\lim_{k\rightarrow0}\omega_{+}=\omega_{\rm fb}.
\end{equation}
Therefore, in this limitation, $\epsilon_{+}\rightarrow1$. $N(0)$ could be calculated that,
\begin{equation}
	N(0)=\lim _{ \rm \epsilon_{ \rm +} \rightarrow 1} \frac{\epsilon_{ \rm {d}}^{2}\left(\epsilon_{ \rm +}^{3}+\epsilon_{ \rm +}\right)}{2 \epsilon_{ \rm {d}}^{2}\left(2 \epsilon_{ \rm +}^{2}-1\right)+2 \epsilon_{ \rm {m}}\left(\epsilon_{ \rm +}^{2}-1\right)^{2}}=1 .
\end{equation}
Then, we have
\begin{equation}
	Ch_{+}=N(\infty)-N(0)=-1,
\end{equation}
which implies that the moiré edge mode has non-trivial topological properties. 

\providecommand{\noopsort}[1]{}\providecommand{\singleletter}[1]{#1}%

\end{document}